\documentclass[sigconf]{acmart}
\usepackage{enumitem}
\setlist[itemize]{leftmargin=*}
\AtBeginDocument{%
  }

\usepackage{color}
\usepackage{diagbox}

\copyrightyear{2023}
\acmYear{2023}
\setcopyright{acmlicensed}
\acmConference[MM '23] {Proceedings of the 31st ACM International Conference on Multimedia}{October 29--November 3, 2023}{Ottawa, ON, Canada.}
\acmBooktitle{Proceedings of the 31st ACM International Conference on Multimedia (MM '23), October 29--November 3, 2023, Ottawa, ON, Canada}
\acmPrice{15.00}
\acmISBN{979-8-4007-0108-5/23/10}
\acmDOI{10.1145/3581783.3612390}

\begin{document}
\title{CLIP-Hand3D: Exploiting 3D Hand Pose Estimation via Context-Aware Prompting}




\author{Shaoxiang Guo}
\affiliation{%
  \institution{Ocean University of China}
  \streetaddress{1299 Sansha Rd}
  \city{Qingdao}
  \country{China}}
\email{guoshaoxiang@stu.ouc.edu.cn}

\author{Qing Cai}
\authornotemark[1]
\affiliation{%
  \institution{Ocean University of China}
  \streetaddress{1299 Sansha Rd}
  \city{Qingdao}
  \country{China}}
\email{cq@ouc.edu.cn}

\author{Lin Qi}
\affiliation{%
  \institution{Ocean University of China}
  \streetaddress{1299 Sansha Rd}
  \city{Qingdao}
  \country{China}}
\email{qilin@ouc.edu.cn}

\author{Junyu Dong}
\authornote{Corresponding author.}
\affiliation{%
  \institution{Ocean University of China}
  \streetaddress{1299 Sansha Rd}
  \city{Qingdao}
  \country{China}}
\email{dongjunyu@ouc.edu.cn}


\begin{abstract}
Contrastive Language-Image Pre-training (CLIP) starts to emerge in many computer vision tasks and has achieved promising performance.
However, it remains underexplored whether CLIP can be generalized to 3D hand pose estimation, as bridging text prompts with pose-aware features presents significant challenges due to the discrete nature of joint positions in 3D space.
In this paper, we make one of the first attempts to propose a novel 3D hand pose estimator from monocular images, dubbed as CLIP-Hand3D, which successfully bridges the gap between text prompts and irregular detailed pose distribution.
In particular, the distribution order of hand joints in various 3D space directions is derived from pose labels, forming corresponding text prompts that are subsequently encoded into text representations.
Simultaneously, 21 hand joints in the 3D space are retrieved, and their spatial distribution (in x, y, and z axes) is encoded to form pose-aware features. 
Subsequently, we maximize semantic consistency for a pair of pose-text features following a CLIP-based contrastive learning paradigm.
Furthermore, a coarse-to-fine mesh regressor is designed, which is capable of effectively querying joint-aware cues from the feature pyramid.
Extensive experiments on several public hand benchmarks show that the proposed model attains a significantly faster inference speed while achieving state-of-the-art performance compared to methods utilizing the similar scale backbone.
Code is available at: \textcolor{magenta}{\url{https://anonymous.4open.science/r/CLIP_Hand_Demo-FD2B/README.md}}.
\end{abstract}
\maketitle
\begin{figure}[htp]
\setlength{\abovecaptionskip}{2pt}
\setlength{\belowcaptionskip}{2pt}
  \centering
  \includegraphics[width=0.9\linewidth]{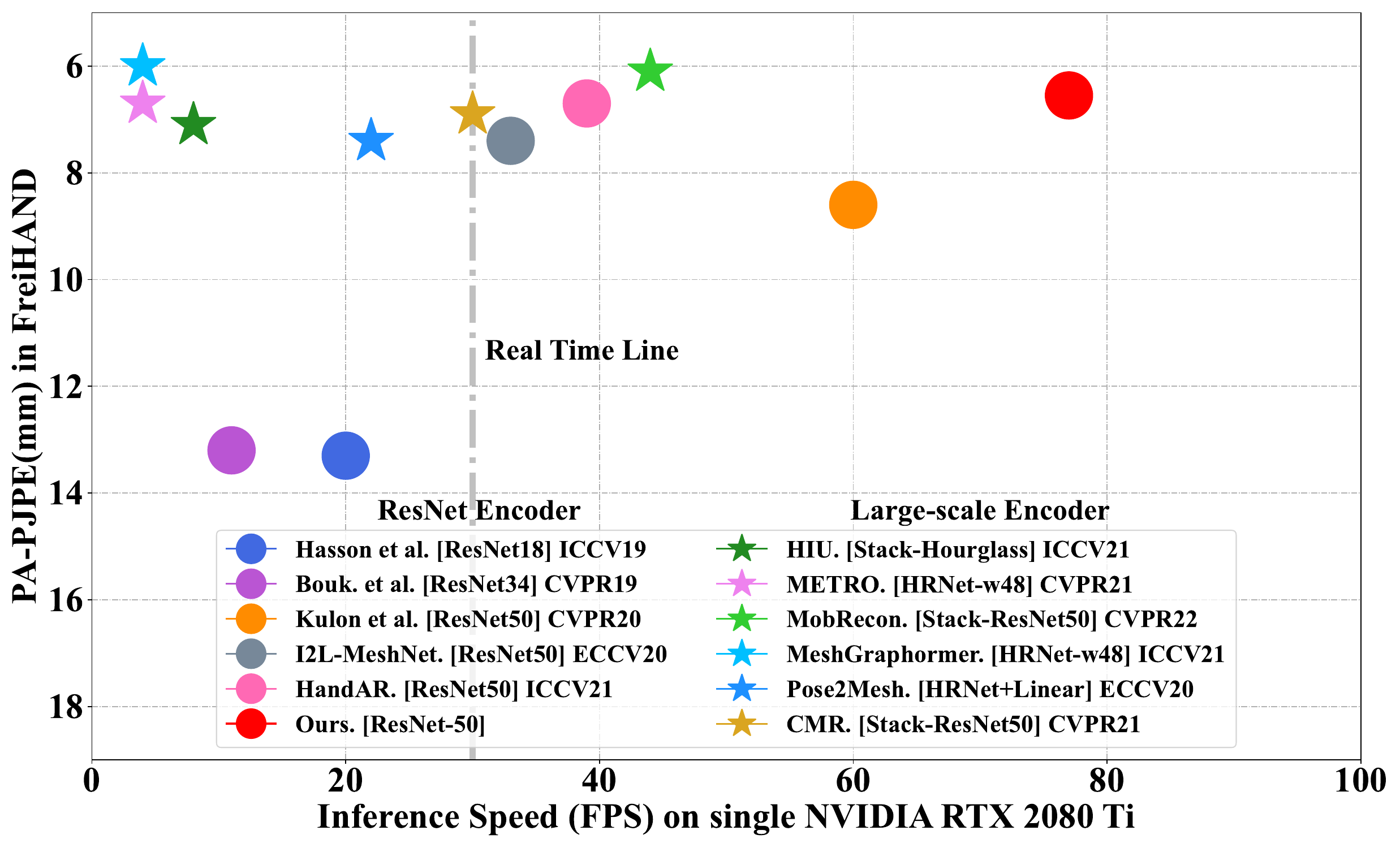}
  \caption{
  Comparison with other methods on the FreiHAND test set. 
  The vertical axis represents PA-PJPE, and the horizontal axis represents inference speed (FPS). 
  The proposed method achieves the \textit{fastest speed} and the \textit{best-performing} compared to methods using the similar-scale visual encoder.
  }
  \label{speed}
\end{figure}
\vspace{-4mm}
\section{Introduction}
3D hand pose estimation, which aims to recover the position of each hand joint from its corresponding monocular image, plays a crucial role in computer vision and behavior understanding due to its extensive application prospects in areas such as virtual reality \cite{tang2021towards}, human-computer interaction \cite{HCI}, gesture recognition \cite{gesture} and sign language translation \cite{signlanguage}. 
Recently, deep convolutional neural network (CNN)-based 3D hand pose estimators have made remarkable progress, with a variety of network architectures being proposed, which can be broadly sorted into two categories: sparse joints regression (i.e., pose recovery \cite{12-Zimmermann-2017, 13-Mueller-2018-GANerated, 22-zhang-2016-3d, 28-cai-2018-weakly-rgb, 30-cai-2019-exploiting-rgb, 20-Doosti-2020-CVPR-hopenet, 29-liu-2019-feature-rgb, 31-guo-2020-graph-rgb, guo-2022-tcsvt-3d, 33-li-2021-exploiting-rgb}) and dense vertices regression (i.e. shape recovery \cite{14-ge-2019-3d, 34-boukhayma-2019-3d-shape, zhang-2021-hand-hiu, lin2021mesh, 19-zhou-2020-monocular, 37-yang-2020-bihand-shape, chen2022mobrecon-CVPR2022, fastmetro, 64-METRO, 60-chen-cvpr21}).
\begin{figure}[htp]
\setlength{\abovecaptionskip}{2pt}
\setlength{\belowcaptionskip}{2pt}
  \centering
  \includegraphics[width=0.9\linewidth]{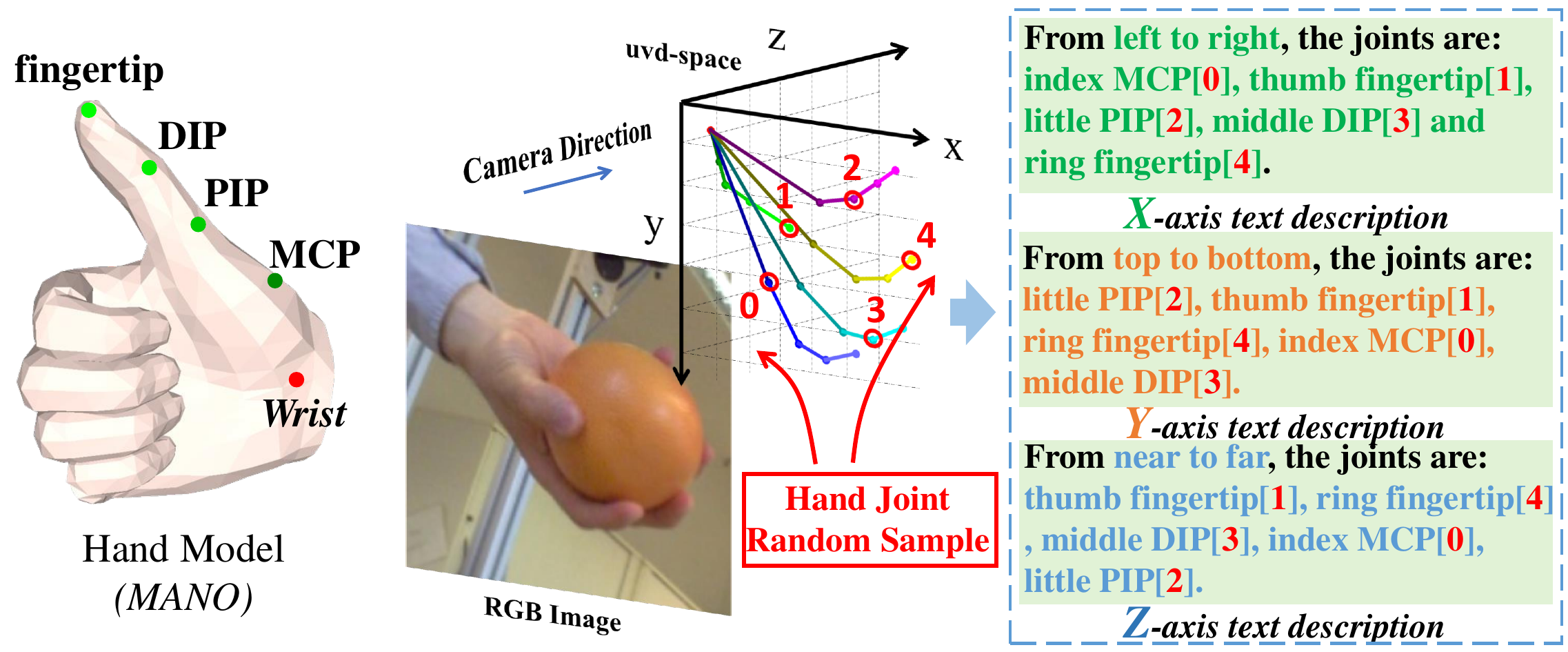}
  \caption{A schematic diagram of text prompts generation. 
  By randomly selecting 5 joints from the 21 joints and generating corresponding text descriptions according to their distribution order in the x, y, and z directions. 
  }
  \label{textprompts}
\end{figure}
However, the success of these methods in achieving reliable 3D hand recovery or understanding mainly depends on large-scale feature encoders, such as 8-Stacked Hourglass \cite{zhang-2021-hand-hiu}, HRNet-w48 \cite{lin2021mesh}, or 2-Stacked ResNet50 \cite{60-chen-cvpr21}, 
which seriously affects the inference speed (as shown in Fig.~\ref{speed}).

To achieve faster inference while maintaining high accuracy, researchers have made two main efforts. 
On the one hand, they carefully design limited-scale visual feature encoders to capture more complex feature details. However, their performance is still unsatisfied due to their inherent defects, that is, they excel at handling and extracting relatively easily recognizable visual semantic representations. 
On the other hand, they pay more attention to expanding datasets. 
Although, this method can achieve superior performance, it is still quite difficult to obtain high-quality datasets with ground truth.
From the above discussions, one question arises immediately:
\textit{Is it possible to leverage high-level human language knowledge to guide them and thereby encode more latent hand semantic details?}
Recently, Radford et al. \cite{radford} introduced the CLIP model to simultaneously input image-text pairs into corresponding feature encoder modules and maximize the feature consistency between them through contrastive learning.
Zhang et al. \cite{zhang2022pointclip}, Xu et al. \cite{groupvit}, Guzhov et al. \cite{audioCLIP} adopted the learning pattern of CLIP and subsequently proposed their respective models.
Experimental results from these studies demonstrate that appropriate text prompts can enrich visual representations, effectively transferring high-level human knowledge into deep neural networks. 
However, unlike image classification and segmentation, 3D hand pose and mesh recovery face challenges in connecting recapitulative text prompts with irregular joint-aware distributions due to the unique nature of labels (discrete 3D joint positions).

To address this issue, we propose a novel and effective model, dubbed as CLIP-Hand3D, as illustrated in Fig.~\ref{textprompts}, which for the first time, successfully transfers discrete 3D joint positions into appropriate text prompts and generates the corresponding text representations.
Specifically, we employ a 1D convolution layer to encode the spatial order of joints along the x, y, and z directions respectively. 
Simultaneously, the shallow perceptual features encoding the Lixel-map are preserved, and subsequently used for matching their corresponding text representation by a contrastive learning paradigm.
In addition, we design a novel hand mesh estimator, incorporating a series of Transformer layers with multi-head self-attention mechanisms. 
Initially, it adopts a coarse-to-fine learning strategy, iteratively refining sparse-to-dense vertex positions with an appropriate positional encoding scheme for the current mesh structure. 
Then, from a global to local perspective, it effectively queries detailed cues from the feature pyramid by utilizing a joint-related feature projection module. 
Furthermore, our model achieves a significant inference speed due to a relatively lightweight visual encoder and a lower-dimensional feature space of the designed regression head. 
As depicted in Fig.~\ref{speed}, our model is capable of real-time inference, achieving a substantially higher FPS value than all other state-of-the-art methods.

The main $\textbf{contributions}$ of our paper are three-folds:
\begin{itemize}
\item A novel model is designed to estimate 3D hand pose and shape from monocular RGB images, exhibiting a \textit{\textbf{significantly faster}} inference speed, while achieving the \textit{\textbf{state-of-the-art}} accuracy compared to methods using the similar scale visual encoders.
\item A novel text feature generation module is designed, which successfully connects irregular joint position labels and text prompts for the first time, thereby achieving consistent matching between pose-aware features and text representations.
\item A novel Transformer mesh regressor is designed, which effectively locates the spatial positional encodings among all sparse-to-dense mesh vertices, thereby matching the inherited joint-related features from the visual encoder.
\end{itemize}
\begin{figure*}[h]
\setlength{\abovecaptionskip}{1pt}
\setlength{\belowcaptionskip}{1pt}
  \includegraphics[width=0.9\textwidth]{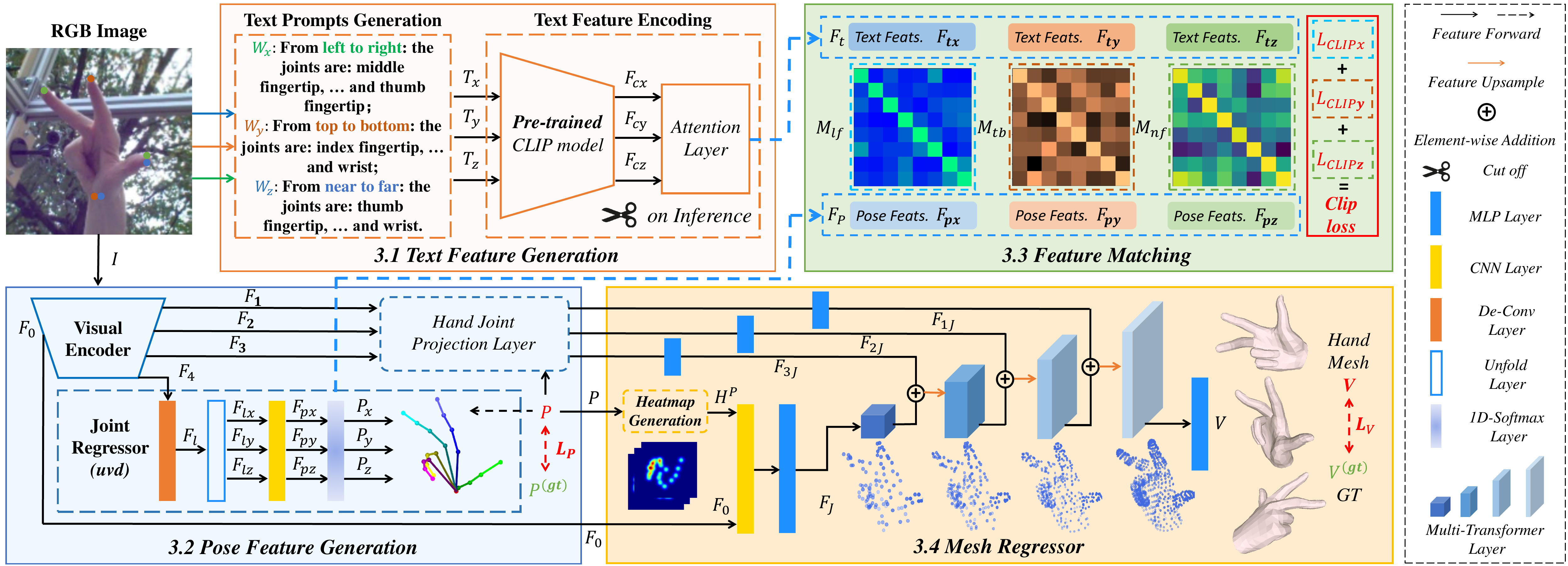}
  \caption{
  A detailed illustration of our proposed pipeline.  
  First, \textit{Text Feature Generation} (see Sec.~\ref{Text Feature Generation} for details) converts 3D pose labels into text prompts and generate text features $F_t$; 
  then, we input image $I$ into a CNN-based \textit{Pose Feature Generation} (see Sec.~\ref{Pose Feature Generation} for details) to extract pose-aware features $F_p$ and regress 3D joint positions $P$; 
  next, we construct matrices $M_{lf}$, $M_{tb}$, and $M_{nf}$ through the \textit{Feature Matching} (see Sec.~\ref{Feature Matching} for details), and maximize the semantic consistency between pose-aware features and their corresponding text representation;
  finally, we estimate reliable 3D hand mesh vertices $V$ through the \textit{Mesh Regressor} (see Sec.~\ref{Mesh Regressor} for details).
  }
  \label{pipeline}
\end{figure*}
\vspace{-2mm}
\section{Related Work}
\textbf{3D Sparse Joints Regression:}
In terms of network structure and regression methods, sparse joints regression works can be divided into several categories: forward kinematics-based regression \cite{park2022handoccnet, zhang-2021-hand-hiu, 44-freihand, 18-iccv-2019-end, 13-Mueller-2018-GANerated, 34-boukhayma-2019-3d-shape}, inverse kinematics-based regression \cite{37-yang-2020-bihand-shape, 19-zhou-2020-monocular}, graph neural network-based pose estimators \cite{31-guo-2020-graph-rgb, 65-lcn-tpami2020, 29-liu-2019-feature-rgb, choi2020pose2mesh, 36-kulon-2020-weakly-shape, 14-ge-2019-3d, chen2022mobrecon-CVPR2022, 60-chen-cvpr21, 30-cai-2019-exploiting-rgb}, 2.5D heatmap-based pose estimators \cite{tang2021towards, 61-iqbal-2018-hand, 75-i2l, 33-li-2021-exploiting-rgb}, and Transformer-based regression networks \cite{64-METRO, lin2021mesh, 63-epipolar-eccv2020, guo-2022-tcsvt-3d, handtransformer, liuhandtransformer}. 
Besides, many researcher 
have explored weakly supervised learning from various perspectives \cite{28-cai-2018-weakly-rgb, zhang-acm-mm-20-weak, spurr2021self-weak, chen2022pseudo, yang2021semihand, tu2023consistent-weak, chen2021model-weak, ohkawa2022domain-weak}.

\textbf{3D Dense Vertices Regression:}
Boukhayma et al. \cite{34-boukhayma-2019-3d-shape} were the first to attempt to estimate 3D hand mesh by predicting MANO model hyperparameters. 
Ge et al. \cite{14-ge-2019-3d} constructed a graph structure for each vertex position of the hand mesh.
Lin et al. successively proposed Transformer-based structures \cite{64-METRO} and a carefully designed Mesh-Graphormer \cite{lin2021mesh} for 3D hand mesh estimation. 
Tang et al. \cite{tang2021towards} presented a model to align estimated hand shape with input image semantics. 
Chen et al. proposed CMR \cite{60-chen-cvpr21} and Mobrcon \cite{chen2022mobrecon-CVPR2022}, which estimate 3D hand mesh in camera space and focus on a mobile-friendly model, respectively.
Moreover, Li et al. \cite{Grpah-Interacting-CVPR2022-two-hand}, Yu et al. \cite{yu2023acr-twohand}, Kim et al. \cite{kim2021end-twohand}, and Lee et al. \cite{lee2023im2hands-twohand} developed a series of impressive models for regressing two hands' 3D pose and shape from monocular images.

\textbf{CLIP-based methods:}
Radford et al. \cite{radford} proposed the CLIP model, which associated textual and visual representations and enabled reliable zero-shot inference. 
Xu et al. \cite{xu2021videoCLIP} introduced the Video-CLIP model, aiming to unify video and textual representations through contrastive learning pretraining. 
Wang et al. \cite{wang2022CLIP-nerf} proposed the first text and image-driven NeRF implementation method. 
Tevet et al. \cite{tevet2022motionCLIP} utilized the knowledge encapsulated in CLIP to introduce a generative model. 
Zhang et al. \cite{zhang2022pointclip} proposed Point-CLIP, achieving alignment between point cloud encoding in CLIP and 3D classification text. 
Xu et al. \cite{groupvit} explored zero-shot transfer from textual supervision to semantic segmentation tasks. 
Rao et al. \cite{rao2021denseCLIP} introduced Dense-CLIP, transforming the original image-text matching problem in CLIP into a pixel-text matching problem.
\textit{Although CLIP-based methods have achieved impressive results in classification and segmentation tasks, no existing work has explored the use of text representations to connect pose-aware features. 
Besides, it is a challenging task to transform irregular joint positions into appropriate text prompts, particularly as it differs from the generalized words used for character class descriptions.}
\vspace{-2mm}
\section{Method}
As discussed earlier, we are dedicated to building a bridge between text prompts and pose-aware features, thereby introducing high-level human knowledge to drive deep neural networks to encode more semantic details.
As depicted in Fig.~\ref{pipeline}, we provide a complete pipeline and network structure containing various modules and gradually introduce the implementation details of each sub-module in the following sections.
\vspace{-2mm}
\subsection{Text Feature Generation}
\label{Text Feature Generation}
\textbf{Text Prompts Generation:}
Unlike tasks such as image classification and segmentation, generating corresponding text prompts from pose labels is not a straightforward process. 
As shown in Fig.~\ref{textprompts}, we propose a method for converting pose labels into text prompts. 
Firstly, we randomly sample $N$ keypoints from the $K=21$ hand keypoints, with ${N}\le{K}$.
Then, we slice the pose label according to the indices of the $N$ keypoints in the $K$ hand keypoints.
For the distribution of $N$ hand keypoints along the $x$-axis, we arrange them in ascending order and generate a set ${N_x}$ describing the sampled points in the $x$-direction according to the index order. 
For a specific hand keypoint i, ${N}_{x^i}$ belongs to ${N_x}$. 
Since there is a high semantic consistency between the image and pose label, a text prefix "From left to right," can be added to describe the order of hand keypoints in the $x$-direction.

Similarly, we obtain the sets ${N_y}$ and ${N_z}$ for the sampled points in the $y$ and $z$ directions, respectively, and generate corresponding description prefixes "From top to bottom," and "From near to far,". 
As shown in Fig.~\ref{textprompts}, assuming \textbf{5} keypoints are selected from the \textbf{21} keypoints, namely index MCP, thumb fingertip, little PIP, middle DIP, and ring fingertip, we can generate three corresponding text prompts \textcolor[RGB]{61,145,64}{$W_x$}, \textcolor[RGB]{210,105,30}{$W_y$}, and \textcolor{blue}{$W_z$}. 
Where \textcolor[RGB]{61,145,64}{$W_x$}: "From \textcolor[RGB]{61,145,64}{left to right}, the joints are index MCP, thumb fingertip, little PIP, middle DIP, and ring fingertip"; \textcolor[RGB]{210,105,30}{$W_y$}: "From \textcolor[RGB]{210,105,30}{top to bottom}, the joints are little PIP, thumb fingertip, ring fingertip, index MCP, and middle DIP"; \textcolor{blue}{$W_z$}: "From \textcolor{blue}{near to far}, the joints are thumb fingertip, ring fingertip, middle DIP, index MCP, and little PIP."

\textbf{Text Feature Encoding:}
Given text prompts $W_x$, $W_y$, and $W_z$, we first employ a similar processing approach to the CLIP \cite{radford} model for tokenization, resulting in $T_x$, $T_y$, and $T_z$. 
Specifically, input text prompts are tokenized into a list of tokens using a mini-batch strategy.
Next, with $T_x$, $T_y$, and $T_z$ provided, we feed them into a \textbf{pre-trained} CLIP model to extract features $F_{cx}$, $F_{cy}$, and $F_{cz}$, and further put them into their corresponding attention layer to get text representations $F_{tx}$, $F_{ty}$, and $F_{tz}$, which describe the spatial order of joints in each dimension.
To adapt the dimensionality of latent joint feature encodings,
and further refine the specific downstream task, we introduce several Transformer modules with multi-head self-attention mechanisms to aid in consistently matching visual features and text representations.
Formally, we have:
\begin{equation}
\setlength{\abovedisplayskip}{2pt}
\setlength{\belowdisplayskip}{2pt}
F_t\Rightarrow
\left\{
\begin{aligned}
    F_{tx}&=\Phi_x(F_c(T_x)),T_x=Token(W_x),   \\
    F_{ty}&=\Phi_y(F_c(T_y)),T_y=Token(W_y),   \\
    F_{tz}&=\Phi_z(F_c(T_z)),T_z=Token(W_z),
\end{aligned}
\right.
\end{equation}
where $\Phi_x$, $\Phi_y$, and $\Phi_z$ represent text encoders based on the Transformer structure; $F_c$ denotes the \textbf{pre-trained} CLIP model; and $Token$ means tokenizing the text prompts.

\begin{figure*}
\setlength{\abovecaptionskip}{0.cm}
\setlength{\belowcaptionskip}{-0.cm}
  \includegraphics[width=0.8\textwidth]{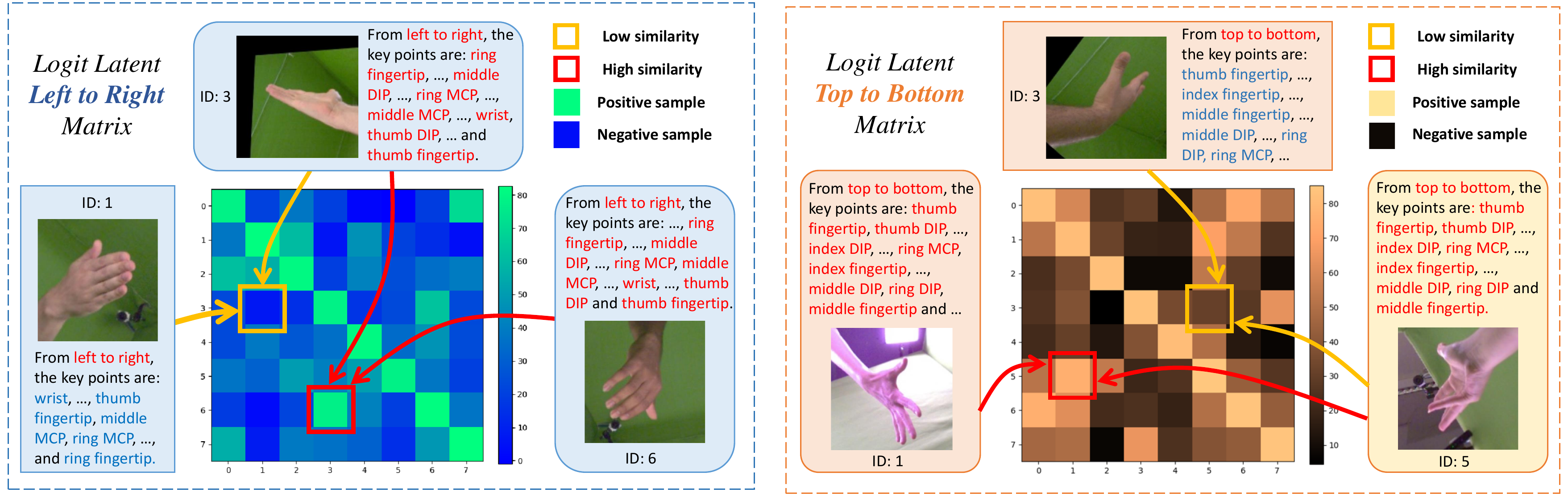}
  \caption{A visualization of the logit latent matrix. 
  The left subplot shows the text-pose feature pairs' matching result in the "left-to-right" direction in one batch; 
  the right subplot shows the feature pairs' matching result in the "top-to-bottom" direction in another batch.
  }
  \label{featurematching}
\end{figure*}
\vspace{-4mm}
\subsection{Pose Feature Generation}
\label{Pose Feature Generation}
\textbf{Visual Encoder:}
Following previous methods
\cite{36-kulon-2020-weakly-shape, tang2021towards, 60-chen-cvpr21}, we employ the original version of ResNet50 \cite{2-ResNet} as the Visual Encoder to encode the input monocular RGB image ${I}\in{{R}^{(224, 224, 3)}}$. 
Given the input monocular image ${I}$, the Visual Encoder yields the following features: 
shallow features $F_0$ and feature pyramids $F_1$, $F_2$, $F_3$, and $F_4$. 
Specifically, $F_0\in{{R}^{(56, 56, 56)}}$ retains the shallow representation, thereby preserving rich high-resolution hand semantics to facilitate refined hand vertices prediction. 
The feature pyramid ($F_1\in{{R}^{(256, 56, 56)}}$, $F_2\in{{R}^{(512, 28, 28)}}$, $F_3\in{{R}^{(1024, 14, 14)}}$) offers a latent distribution from global to local, which corresponds to the coarse-to-fine mesh feature sampling in the Mesh Regressor module. 
$F_4\in{{R}^{(2048, 8, 8)}}$ is passed to the Joint Regressor module to output joint spatial positions in the x, y, and z directions by feature upsampling (de-convolution layer).

\textbf{Joint Regressor:}
Given the feature encoding $F_4$ produced by the Visual Encoder, the Joint Regressor module generates spatial positions of hand joints (in \textit{uvd} space). 
Firstly, we employ a deconvolution layer (kernel-size=4, stride=2) to perform feature upsampling on $F_4$, obtaining the feature $F_l$, and subsequently reducing its number of channels from 1024 to 256.
Following the Lixel-Map-based approach proposed by \cite{75-i2l, tang2021towards}, we unfold the feature $F_l$ along the specified dimension to obtain feature $F_{lx}, F_{ly}, F_{lz}$, and further apply 1D convolution to obtain the latent features $F_{p}:{\{F_{px}, F_{py}, F_{pz}\}}$.
Then, we design corresponding 1D softmax layers according to the heatmap size to capture the maximum response points of hand joints $P:\{P_x, P_y, P_z\}$ in the x, y, and z directions, respectively. 
By identifying the position indices of the maximum response points in the feature map and using the concatenate operation, the Joint \textit{uvd} Regressor module outputs the 3D hand joint positions $P\in{R^{(21, {3})}}$ in \textit{uvd} space.
Formally, we have:
\begin{equation}
\setlength\abovedisplayskip{2pt}
\setlength\belowdisplayskip{2pt}
P\Rightarrow
\left\{
\begin{aligned}
    P_x&=Argmax(Soft(F_{px})),F_{px}=Conv1d(F_{lx}), \\
    P_y&=Argmax(Soft(F_{py})),F_{py}=Conv1d(F_{ly}), \\
    P_z&=Argmax(Soft(F_{pz})),F_{pz}=Conv1d^{*}(F_{lz}),
\end{aligned}
\right.
\end{equation}
where $Conv1d$ means 1D-Convolution layer; $*$ denotes multi-layer structure; $Soft$ representes 1D-Softmax activation layer; $Argmax$ means the index of the maximum value.
\vspace{-2mm}
\subsection{Feature Matching}
\label{Feature Matching}
Building on several prior CLIP-based works, we pass images and texts through their respective feature encoders to obtain pose-aware features $F_p$ and text representations $F_t$. 
To enrich the pose-aware features using text representations, we together project $F_p$ and $F_t$ into a common feature embedding space and compute the feature similarity between them.
Employing a mini-batch optimization strategy with a batch size of \textbf{B}, we construct the logit latent matrix \textbf{M} of shape [3, B, B], treating all matched pose-text pairs as positive samples and the remaining non-matching pose-text pairs as negative samples. 
Intuitively, all values on the diagonal of the logit latent matrix represent the B pose-text pairs in the current batch.
Formally, we have:
\begin{equation}
\setlength\abovedisplayskip{2pt}
\setlength\belowdisplayskip{2pt}
\left\{
\begin{aligned}
    M_{lr}&=\tau_{x}\hat{F}_{px}\cdot\hat{F}_{tx}^T,   \\
    M_{tb}&=\tau_{y}\hat{F}_{py}\cdot\hat{F}_{ty}^T,   \\
    M_{nf}&=\tau_{z}\hat{F}_{pz}\cdot\hat{F}_{tz}^T,
\end{aligned}
\right.
\end{equation}
where $M_{lr}$, $M_{tb}$, and $M_{nf}$ represent three different logit latent matrices and $M=\{M_{lr}, M_{tb}, M_{nf}\}$; 
$\tau_{x}$, $\tau_{y}$, and $\tau_{z}$ denote the corresponding learnable temperature parameters to scale the logit matrix respectively; 
$\hat{F}_*$ stands for visual or text feature representations after L2 regularization; 
$\cdot$ represents matrix multiplication; and ${T}$ indicates matrix transpose.

The left sub-figure in Fig.~\ref{featurematching} presents a matching result of visual and text representations in the "horizontal" direction for a batch containing 8 samples, where the \textcolor[RGB]{61,145,64}{green} elements on the diagonal represent positive samples and the \textcolor{blue}{blue} elements denote negative samples. 
Specifically, we use \textcolor[RGB]{238,180,34}{yellow} and \textcolor{red}{red} boxes to display two image-text matching pairs for detailed illustration.
On one hand, images with ID 3 and ID 6 have a high matching similarity.
This is because their hand joint distributions exhibit high consistency when viewed from the "From left to right" perspective ("ring fingertip, middle DIP, ring MCP, middle MCP, wrist, thumb DIP, and thumb fingertip").
We highlight their corresponding text prompts in \textcolor{red}{red}. 
On the other hand, images with ID 1 and ID 3 have a lower pose-aware features similarity.

The right sub-figure in Fig.~\ref{featurematching} shows a matching result of visual and text representations in the "vertical" direction for another batch containing 8 samples, where the \textcolor[RGB]{238,180,34}{yellow} elements on the diagonal represent positive samples, and the \textbf{\textcolor{black}{black}} elements denote negative samples. 
Intuitively, image samples with ID 1 and ID 5 have high similarity in pose distribution from top to bottom, while samples with ID 3 and ID 5 have low similarity.
The hands in images with ID 1 and ID 5 are both pointing "downward" while the hand in the image with ID 3 appears to be pointing "upward".
\subsection{Mesh Regressor}
\label{Mesh Regressor}
To forward the Joint \textit{uvd} into the Mesh Regressor module, we first generate the corresponding heatmap $H^{p}$ according to the predefined heatmap size. 
Then, we concatenate the $F_0$ encoding the shallow semantic information of the hand with heatmap $H^{p}$ along the specified dimension, and after using the CNN and MLP layer to adjust the number of channels and feature dimensions.
Finally, we obtain the 3D hand shape feature encoding $F_{J}$.

\textbf{Mesh Regressor Layer}:
As depicted in Fig.~\ref{pipeline}, we designed a mesh feature refinement network with a coarse-to-fine approach, implemented in the order of [256, 128, 64, 32] to refine mesh features. 
Additionally, we devised a mesh node sampling network with a sparse-to-dense progression, increasing the number of hand mesh points in the order of [21, 98, 389, 778]. 
We employed the upsampling network based on GNN to expand the number of nodes while utilizing an MLP layer to adjust the feature dimensions. 
Following each feature dimension adjustment, we introduce a Transformer-based multi-head attention regression network to enhance the mesh features and enable self-attention interaction at that layer. 
Notably, the Transformer-based regressor maintains the original node's positional encoding, as it does not alter the number of nodes or feature dimensions at each layer. 
Finally, we designed an MLP layer for regressing and predicting the spatial coordinates of the mesh $V$.

\textbf{Feature Pyramid Projection}:
To fully harness the hand feature encoding in the Visual Encoder, we designed a Joint-based Feature Pyramid Projection module. 
As illustrated in Fig.~\ref{pipeline}, we project the features of each level in the Feature Pyramid according to the Joint \textit{uvd} position. 
Specifically, we adopt a structure similar to U-Net \cite{unet}, linking the mesh feature encodings at each level with the corresponding visual representation, ensuring the consistency of feature encoding levels. 
For instance, when considering the $F_1$ in the feature pyramid, we project the joint \textit{uvd} position to obtain the corresponding feature sampling $F_{1J}$, and then pass it to the feature encoding representing the dense hand mesh. 
High-resolution feature maps possess a smaller spatial receptive field, which aids in preserving more shallow hand semantics to assist the corresponding hand mesh feature encoding in accurately determining the hand mesh point positions.
\vspace{-4mm}
\subsection{Loss Functions}
We mainly applied the following loss functions: 

\textbf{1) Supervised Learning Loss}:
For 3D hand pose and shape recovery from monocular RGB image, we employ $L1$ norm for 3D hand joint loss $\mathcal{L}_{P}$ and vertices loss $\mathcal{L}_{V}$.
Formally, we have:
\begin{equation}
\setlength\abovecaptionskip{2pt}
\setlength\belowcaptionskip{2pt}
\mathcal{L}_{P}=\sum_{i=1}^{J}{\big{|}}{\big{|}}{P}_{i}^{(gt)}-{P}_{i}{\big{|}}{\big{|}}_1,
\mathcal{L}_{V}=\sum_{i=1}^{K}{\big{|}}{\big{|}}{V}_{i}^{(gt)}-{V}_{i}{\big{|}}{\big{|}}_1,
\end{equation}
where ${P}_{i}$ and ${P}_{i}^{(gt)}$ represent the predicted 3D hand pose and its ground truth, respectively; ${V}_{i}$ and ${V}_{i}^{(gt)}$ represent the predicted hand vertices and their ground truth, respectively.

\textbf{2) Norm Loss}:
To maintain the stability of predicted hand mesh surface normals and vertices during the training process, we applied normal loss $\mathcal{L}_N$ and vertex loss $\mathcal{L}_E$.
\begin{equation}
\setlength\abovedisplayskip{2pt}
\setlength\belowdisplayskip{2pt}
\begin{aligned}
\mathcal{L}_{N}&=\sum_{{c}\in{C}}\sum_{(i, j)\subset{C}}\bigg{|}{\frac{V_i-V_j}{||V_i-V_j||_2}}\cdot{n}_c^{(gt)}\bigg{|}_1,
\\
\mathcal{L}_{E}&=\sum_{{c}\in{C}}\sum_{(i, j)\subset{C}}\big{|}||V_i-V_j||_2 -||V_i^{gt}-V_j^{gt}||_2\big{|}_1,
\end{aligned}
\end{equation}
where $V$ and $C$ represent the predicted vertices and its corresponding triangular faces respectively, 
$n_{c}^{(gt)}$ denotes unit normal vector for each face ${c}\in{C}$, $V^{(gt)}$ means hand vertices ground-truth.

\textbf{3) Consistency Loss}:
Following several self-supervision related works, we applied 2D and 3D consistency losses ($\mathcal{L}_{C2d}$ and $\mathcal{L}_{C3d}$) to supervise the predicted pose and vertices.
\begin{equation}
\setlength\abovedisplayskip{2pt}
\setlength\belowdisplayskip{2pt}
\mathcal{L}_{C2d}={\big{|}}{\big{|}}Aff(\hat{P}_1)-\hat{P}_2{\big{|}}{\big{|}}_1,
\mathcal{L}_{C2d}={\big{|}}{\big{|}}Rot(V_1)-V_2{\big{|}}{\big{|}}_1,
\end{equation}
where $V_1$, $\hat{P}_1$ and $V_2$, $\hat{P}_2$ represent the predicted hand vertices and joint \textit{uvd} positions under different viewpoints, and $Rot$ and $Aff$ denote the rotation and affine transformation matrix.

\textbf{4) CLIP Loss}:
To maximize the values distributed along the diagonal of the logit matrix, thereby achieving a match between text and pose-aware features, we introduce CLIP loss $\mathcal{L}_{CLIP}$ to supervise a batch containing B image-text pairs.
Formally, we have:
\begin{equation}
\setlength\abovedisplayskip{2pt}
\setlength\belowdisplayskip{2pt}
\begin{aligned}
\mathcal{L}_{CLIP*}&=\bigg{(}-\frac{1}{B}\sum_{i=1}^{B}log\frac{exp(\hat{F}_{p*}\cdot\hat{F}_{t*}/\tau_{*})}{\sum_{j=1}^{B}exp(\hat{F}_{p*}\cdot\hat{F}_{t*}/\tau_{*})}\bigg{)}
\\
&+\bigg{(}-\frac{1}{B}\sum_{i=1}^{B}log\frac{exp(\hat{F}_{t*}\cdot\hat{F}_{p*}/\tau_{*})}{\sum_{j=1}^{B}exp(\hat{F}_{t*}\cdot\hat{F}_{p*}/\tau_{*})}\bigg{)},
\end{aligned}
\end{equation}
where $\mathcal{L}_{CLIP*}$ describes the consistency loss of pose-text pairs along the current coordinate axis direction; 
$\hat{F}_{l*}$ and $\hat{F}_{t*}$ represent the corresponding normalized pose representation and text feature encoding, respectively; 
Subsequently, we add the CLIP losses in three directions (x, y, z) and compute the average to obtain $\mathcal{L}_{CLIP}=\frac{1}{len(S)}\sum_{*\in{S}}\mathcal{L}_{CLIP*},S=\{x, y, z\}$.

Finally, we train the whole framework to optimize all the learnable parameters in an \textbf{end-to-end} manner.
Formally, we have:
\begin{equation}
        \mathcal{L}={\alpha_1}(\mathcal{L}_{P} + \mathcal{L}_{V})+{\alpha_2}(\mathcal{L}_{N} + \mathcal{L}_{E})
        + {\alpha_3}(\mathcal{L}_{C2d} + \mathcal{L}_{C3d}) + {\alpha_4}\mathcal{L}_{CLIP},
\end{equation}
where the hyper-parameters ${\alpha_1}$, ${\alpha_2}$, ${\alpha_3}$ and ${\alpha_4}$ are balance factors to weight the losses, where ${\alpha_1}$ = 1.0, ${\alpha_2}$ = 0.05, ${\alpha_3}$ = 0.1 and ${\alpha_4}$ = 0.1.
\begin{table*}[]
\caption{Quantitative evaluation. 
Comparison with other SOTA methods on the FreiHAND \cite{44-freihand} test set. 
We evaluated them by PA-PJPE, PA-PVPE, F@5/15mm and FPS metrics.
We use bold font to indicate the best performance, and use "$\_$" to represent the second-best performance.} 
\footnotesize
\centering
\setlength{\tabcolsep}{0.75mm}{%
\begin{tabular}{@{}ccccccc@{}}
\toprule
Method                  & Backbone        & PA-PJPE $\downarrow$ & PA-PVPE $\downarrow$ & F@5mm $\uparrow$ & F@15mm $\uparrow$ & \textcolor[RGB]{65,105,225}{\textbf{FPS}} $\uparrow$ \\ \midrule
MANO CNN (ICCV 19) \cite{44-freihand}       & ResNet50        & 10.9    & 11.0    & 0.516 & 0.934  & -   \\
Hasson et al. (CVPR 19) \cite{56-obman}  & ResNet18        & 13.3    & 13.3    & 0.429 & 0.907  & 20  \\
Boukh.et al. (CVPR 19) \cite{34-boukhayma-2019-3d-shape}   & ResNet34        & 13.2    & 35.0    & 0.427 & 0.894  & 11  \\
Kulon et al. (CVPR 20) \cite{36-kulon-2020-weakly-shape}   & ResNet50        & \textcolor[RGB]{65,105,225}{8.4}     & \textcolor[RGB]{65,105,225}{8.6}     & \textcolor[RGB]{65,105,225}{0.614} & \textcolor[RGB]{65,105,225}{0.966}  & \textcolor[RGB]{65,105,225}{\underline{60}}  \\
I2L-MeshNet (ECCV 20) \cite{75-i2l}    & ResNet50        & 7.4     & 7.6     & 0.681 & 0.973  & 33  \\
I2UV-HandNet (ICCV 21) \cite{i2uvhandnet}   & ResNet50        & 7.2     & 7.4     & 0.682 & 0.973  & -   \\
HandAR (ICCV 21) \cite{tang2021towards}         & ResNet50        & \textcolor[RGB]{65,105,225}{\underline{6.7}}     & \textcolor[RGB]{65,105,225}{\textbf{6.7}}     & \textcolor[RGB]{65,105,225}{\underline{0.724}} & \textcolor[RGB]{65,105,225}{\underline{0.981}}  & \textcolor[RGB]{65,105,225}{39}  \\
CycleHand (ACM MM 22) \cite{acm-22-gao2022cyclehand}    & ResNet50        & 8.3     & 8.3     & 0.631 & 0.967  & -   \\ \midrule
Ours.                   & ResNet50        & \textcolor[RGB]{65,105,225}{\textbf{6.6}}     & \textcolor[RGB]{65,105,225}{\textbf{6.7}}     & \textcolor[RGB]{65,105,225}{\textbf{0.728}} & \textcolor[RGB]{65,105,225}{\textbf{0.981}}  & \textcolor[RGB]{65,105,225}{\textbf{77}}  \\ \midrule
Pose2Mesh (ECCV 20) \cite{choi2020pose2mesh}      & HRNet+Linear          & 7.4     & 7.6     & 0.683 & 0.973  & 22  \\ 
MANO GCN (ICME 21) \cite{manogcn}       & HRNet-w48       & 9.5     & 9.5     & 0.579 & 0.950  & -   \\
CMR (CVPR 21) \cite{60-chen-cvpr21}            & Stack-ResNet50  & 6.9     & 7.0     & 0.715 & 0.977  & \underline{30}  \\
METRO (ICCV 21) \cite{64-METRO}          & HRNet-w48       & 6.7     & 6.8     & 0.717 & 0.981  & 4   \\
HIU (ICCV 21) \cite{zhang-2021-hand-hiu}            & Stack-Hourglass & 7.1     & 7.3     & 0.699 & 0.974  & 9   \\
MobRecon (CVPR 22) \cite{chen2022mobrecon-CVPR2022}       & Stack-ResNet50  & \underline{6.1} & \underline{6.2}     & \underline{0.760} & \underline{0.984}  & \textbf{45}  \\
Fast-METRO (ECCV 22) \cite{fastmetro}     & HRNet-w48       & 6.5     & -       & -     & 0.982  & 14  \\
MeshGraphormer (ICCV 21) \cite{lin2021mesh} & HRNet-w48       & \textbf{5.9}     & \textbf{6.0}     & \textbf{0.765} & \textbf{0.987}  & 4  
\\ 
\bottomrule
\end{tabular}%
}
\label{freihand}
\end{table*}
\vspace{-2mm}
\section{Experiments}
\subsection{Datasets \& Metrics}
\textbf{FreiHAND}:The FreiHAND dataset (FHD) \cite{44-freihand} contains \textbf{130,240} training images from 32 characters of different genders and ethnic backgrounds, holding either nothing or various standard daily necessities. 
The test set in this dataset includes \textbf{3,960} samples collected from specific outdoor and office scenes. 

\textbf{RHD}:The RHD \cite{12-Zimmermann-2017} dataset is a synthetic dataset, which poses a significant challenge due to the complex texture-based backgrounds, rich hand gestures, and severe self-occlusion present in the hand objects. 
Following the same settings with the previous methods, we used \textbf{41,258} images for training and \textbf{2,728} images for testing.

\textbf{STB}:The Stereo Hand Pose Tracking Benchmark (STB) \cite{22-zhang-2016-3d} is a real-world dataset. 
Following the same settings with the previous methods, we use STB-SK subset includes \textbf{15,000} RGB images for training and another \textbf{3,000} RGB images for testing, all of which provide accurate hand keypoint annotations.

\textbf{Real-world}:The real-world dataset \cite{14-ge-2019-3d} consists of two parts, the training set and the test set. 
The training set contains over \textbf{300,000} synthesized hand images and corresponding labels, while the test set provides more than \textbf{500} real-world hand images.

We use the following metrics to quantitatively evaluate model performance: 
\textbf{PJPE} (per joint position error), 
\textbf{PVPE} (per vertice position error), 
\textbf{PA-PJPE},
\textbf{PA-PVPE},
\textbf{Median PJPE}, 
\textbf{3D PCK}, 
\textbf{AUC} (area under PCK curve), 
\textbf{F@5mm}, and \textbf{F@15mm}.
\vspace{-2mm}
\subsection{Implementation Details}
During the training and inference stages, we use \textbf{PyTorch} as the framework to conduct all experiments. 
We train our full model on a single NVIDIA RTX 3090 and a single NVIDIA RTX 2080Ti for image inference.
Initially, we pre-train the weight parameters of the Visual Encoder (ResNet50) and Joint \textit{uvd} Regressor. 
Before this, we load the ImageNet pre-trained weight parameters into the Visual Encoder.
Subsequently, we fine-tune the entire network parameters in an \textbf{end-to-end} optimization manner.
The specific process involves using the AdamW optimizer \cite{adamw} with a mini-batch size of 48 and training for 200 epochs.
The initial learning rate is set at 1e-3, and the learning rate schedule follows a fixed-step decay strategy, where the learning rate is reduced to 0.25 times the previous rate every 50 epochs. 
Although our proposed model contains several MLP layers, the feature dimensions of the designed fully connected layers are relatively low (e.g., 128, 64, 32), which lays the foundation for fast inference. 
On a single NVIDIA RTX 2080Ti, the inference speed surpasses \textbf{77 FPS}; 
on a single NVIDIA RTX 3090, the inference speed achieves \textbf{120 FPS}.
\begin{table}[ht]
\caption{Quantitative evaluation. 
Comparison with other SOTA methods on the RHD and STB test sets. 
We use bold font to indicate the best performance, and use "$\_$" to represent the second-best performance.
}
\footnotesize
\centering
\setlength{\tabcolsep}{0.45mm}{%
\begin{tabular}{@{}ccccc@{}}
\toprule
Dataset   & \multicolumn{2}{c}{STB \cite{22-zhang-2016-3d}}    & \multicolumn{2}{c}{RHD \cite{12-Zimmermann-2017}} \\ \midrule
\begin{tabular}[c]{@{}c@{}}Evaluation\end{tabular} & \begin{tabular}[c]{@{}c@{}}AUC $\uparrow$ \end{tabular} & \begin{tabular}[c]{@{}c@{}}PJPE$\downarrow$\end{tabular} & \begin{tabular}[c]{@{}c@{}}AUC $\uparrow$\end{tabular} & \begin{tabular}[c]{@{}c@{}}PJPE$\downarrow$\end{tabular} 
\\ \midrule
Zimmer. et al. (CVPR 2017) \cite{12-Zimmermann-2017}  & 0.948   & -   & 0.670  & 30.42  \\
Spurr et al. (CVPR 2018) \cite{57-spurr} & 0.983  & 8.56  & 0.849  & 19.73   \\
Cai et al. (CVPR 2018) \cite{28-cai-2018-weakly-rgb}  & 0.994  & - & 0.887  & -  \\
Ge et al. (CVPR 2019) \cite{14-ge-2019-3d} & \textcolor[RGB]{65,105,225}{0.998}  & \textcolor[RGB]{65,105,225}{6.37} & 0.920  & - \\
Boukh. et al. (CVPR 2019) \cite{34-boukhayma-2019-3d-shape} & -  & 9.76 & -  & -  \\
Zhang et al. (ICCV 2019) \cite{18-iccv-2019-end}  & 0.995  & - & 0.901  & - \\
Yang et al. (CVPR 2019) \cite{45-yang2019aligning} & 0.996 & 7.05 & 0.943 & 13.14 \\
Zhou et al. (CVPR 2020) \cite{19-zhou-2020-monocular} & 0.898 & -  & 0.856 & -   \\
Wu et al. (ACM MM 2020) \cite{wu-mm-hand3d}  & \textcolor[RGB]{65,105,225}{\underline{0.999}} & -  & 0.929  & -  \\
Kulon et al. (CVPR 2020) \cite{36-kulon-2020-weakly-shape} & - & -   & 0.956   & 10.92  \\
Yang et al. (BMVC 2020) \cite{37-yang-2020-bihand-shape}  & 0.997  & 10.05  & 0.951  & 12.76   \\
Cai et al. (TPAMI 2020) \cite{51-TPAMI-cai20203d}  & 0.996  & 7.10   & 0.915   & -  \\
Zhang et al. (ACM MM 2020) \cite{zhang-acm-mm-20-weak}   & 0.996  & -   & -    & -  \\
Li et al. (AAAI 2021) \cite{33-li-2021-exploiting-rgb}   & 0.996  & -    & \textcolor[RGB]{65,105,225}{0.960}  &\textcolor[RGB]{65,105,225}{10.65} \\    
Chen et al. (CVPR 2021) \cite{60-chen-cvpr21}   & -  & -    & 0.949  & -   \\
Zhang et al. (ICCV 2021) \cite{zhang-2021-hand-hiu} & 0.995 & -  & \textcolor[RGB]{65,105,225}{\underline{0.964}} & - \\ 
CycleHand (ACM MM 2022) \cite{acm-22-gao2022cyclehand}  & -  & 7.94  & -  & -  \\ \midrule
Ours  & \textcolor[RGB]{65,105,225}{\textbf{0.999}}  & \textcolor[RGB]{65,105,225}{\textbf{6.35}}  & \textcolor[RGB]{65,105,225}{\textbf{0.965}}  & \textcolor[RGB]{65,105,225}{\textbf{10.58}}  \\ \bottomrule
\end{tabular}%
}
\label{stbrhd}
\end{table}
\begin{figure*}
\setlength{\abovecaptionskip}{2pt}
\setlength{\belowcaptionskip}{2pt}
  \includegraphics[width=0.7\textwidth]{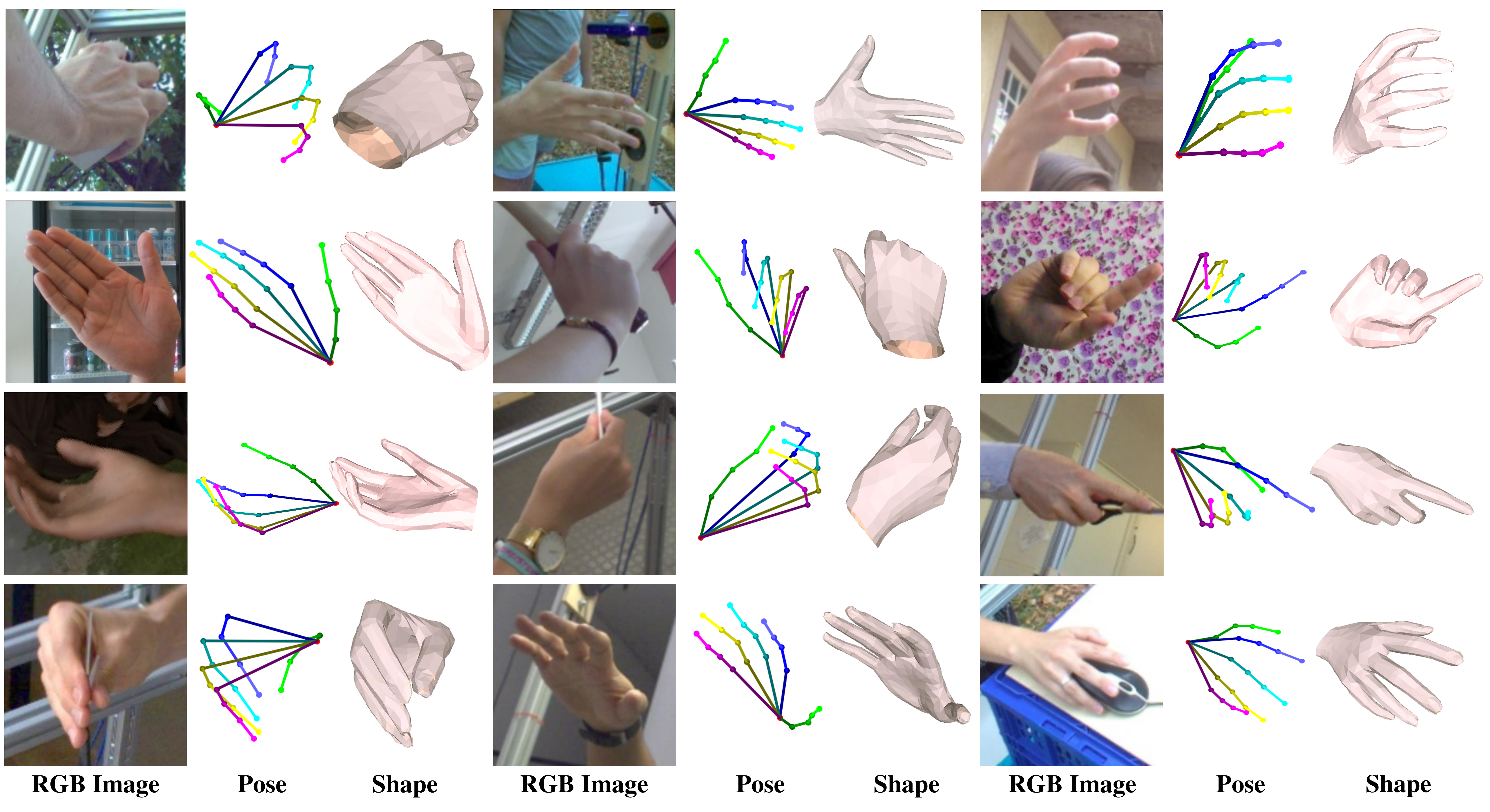}
  \caption{Qualitative evaluation results. 
  We randomly selected some samples from the testset of FreiHAND \cite{44-freihand}, STB \cite{22-zhang-2016-3d}, RHD \cite{12-Zimmermann-2017} and real-world \cite{14-ge-2019-3d} datasets, and showed their 3D hand pose and shape in one camera view.}
  \label{Qualitative}
\end{figure*}
\begin{figure*}
\setlength{\abovecaptionskip}{2pt}
\setlength{\belowcaptionskip}{2pt}
  \includegraphics[width=0.65\textwidth]{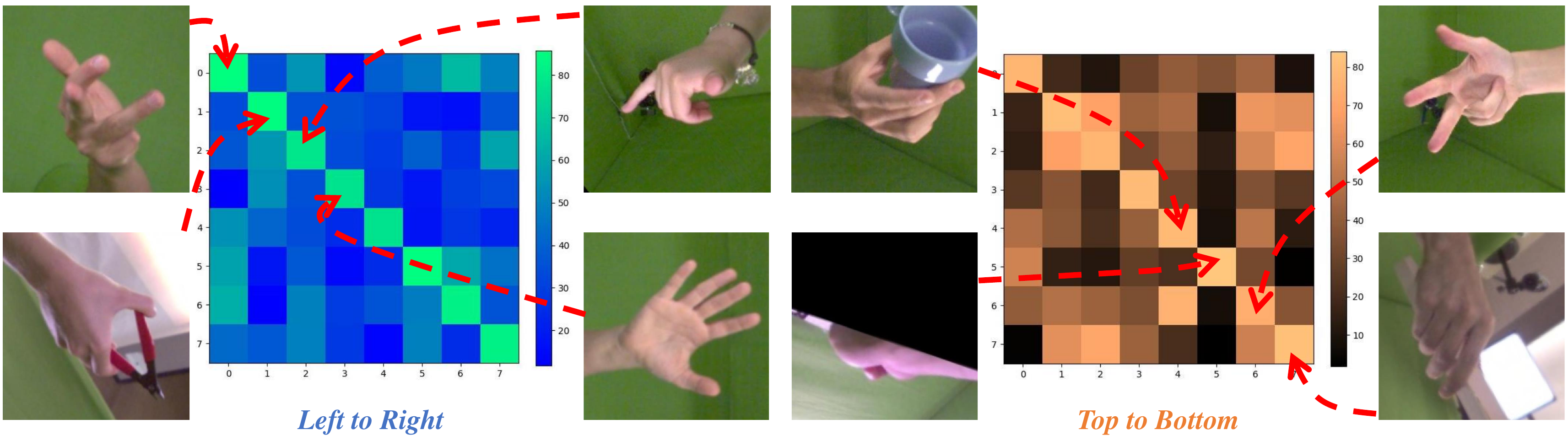}
  \caption{Qualitative evaluation results. 
  A mini-batch with 8 images: assessing textual descriptions via logits latent matrices "Left to Right" and "Top to Bottom".}
  \label{batchvis}
\end{figure*}
\vspace{-2mm}
\subsection{Quantitative and Qualitative Results}
\subsubsection{Quantitative Results}
For the FreiHAND dataset, we primarily conduct a detailed comparison with the current state-of-the-art (SOTA) methods. 
As shown in Table.~\ref{freihand}, among methods that employ the original ResNet50 as the visual encoder (MANO CNN \cite{44-freihand}, Kulon et al. \cite{36-kulon-2020-weakly-shape}, I2L-MeshNet \cite{75-i2l}, I2UV-HandNet \cite{i2uvhandnet}, HandAR \cite{tang2021towards}, and CycleHand \cite{acm-22-gao2022cyclehand}), our approach achieves SOTA performance and inference speed (compared to the \textcolor[RGB]{65,105,225}{\textit{fastest-performing}} method by Kulon et al., F@5: 0.614 vs. \textbf{0.728}, FPS: 60 vs. \textbf{77}; compared to the \textcolor[RGB]{65,105,225}{\textit{best-performing}} method by HandAR, F@5: 0.724 vs. \textbf{0.728}, FPS: 39 vs. \textbf{77}).
It can also be found that, when compared with SOTA models with large-scale visual encoders (MeshGraphormer \cite{lin2021mesh}, and MobRecon \cite{chen2022mobrecon-CVPR2022}), our method can still achieve speed compaction (FPS: 4, 45 vs. \textbf{77}) with competitive performance (6.1, 5.9 vs. \textbf{6.6}mm)
Visual encoders with relatively large complexity (Stack-ResNet50 \cite{60-chen-cvpr21, chen2022mobrecon-CVPR2022}, HRNet-w48 \cite{64-METRO, lin2021mesh, fastmetro, manogcn}, N-Stacked Hourglass \cite{zhang-2021-hand-hiu, guo-2022-tcsvt-3d}) can encode more semantic details of hand image yet can bring trouble for the faster network inference. 
Besides, as the scale of the visual encoder model increases, our model can also achieve the accuracy of state-of-the-art performance.
Note that some methods' performance using supplementary or mixed datasets is not included in our statistics.

In addition, to further validate the superior performance of the proposed method, as shown in Table.~\ref{stbrhd}, we compared the model comparisons on the STB and RHD datasets with several other methods. 
The STB dataset contains a relatively limited distribution of hand poses, making it easier to fit. 
In the 20-50mm AUC curve evaluation, our method achieved SOTA performance (AUC: \textbf{0.999}, PJPE: \textbf{6.35mm}).
On the other hand, the RHD dataset is a synthetic dataset, with samples featuring highly complex textures and severe self-occlusions. 
Compared to the currently the \textcolor[RGB]{65,105,225}{\textit{best-performing}} methods proposed by Zhang et al. \cite{zhang-2021-hand-hiu} and Li et al. \cite{33-li-2021-exploiting-rgb}, our proposed model achieved the highest AUC value (0.960, 0.964 vs. \textbf{0.965}) and the lowest PJPE (10.65mm vs. \textbf{10.58}mm). 
For the method proposed by Kulon et al. \cite{36-kulon-2020-weakly-shape}, which uses the same visual encoder, our model still has a certain advantage (FPS: 60 vs. \textbf{77}, AUC: 0.956 vs. \textbf{0.965}, PJPE: 10.92 vs. \textbf{10.58}).
Note that the methods that use supplementary datasets were not included in the statistics.
\begin{table}[htp]
\caption{Ablation Study.
Comparison model performance on the FreiHAND validation set under different structures.
}
\footnotesize
\setlength{\tabcolsep}{0.5mm}{%
\begin{tabular}{@{}ccccc@{}}
\toprule
\diagbox{Structures}{Matrices}                       & AUC(0-30mm)    & PJPE           & Median PJPE  & PA-PJPE        \\ \midrule
baseline               & 0.758          & 15.46          & 12.78        & 7.47          \\
w/o CLIP               & 0.763          & 14.86          & 12.27        & 7.29          \\
w/o Feature Projection & 0.769          & 14.56          & 12.01        & 7.11          \\
w/o Sparse-to-dense    & 0.773          & 14.21          & 11.80        & 6.97          \\
Full model.            & \textbf{0.776} & \textbf{13.95} & \textbf{11.58} & \textbf{6.88} \\ \bottomrule
\end{tabular}
}
\label{ablation1}
\end{table}
\subsubsection{Qualitative Results}
As illustrated in Fig.~\ref{Qualitative}, we randomly selected over a dozen test images from the FreiHAND \cite{44-freihand}, STB \cite{22-zhang-2016-3d}, RHD \cite{12-Zimmermann-2017} and real-world \cite{14-ge-2019-3d} datasets and employed the proposed method to infer reliable 3D hand poses and shapes. 
These images exhibit significant variations in hand poses, different degrees of self-occlusion, and even some hand-object interactions. 
Intuitively, our model can estimate reliable 3D hand poses and corresponding mesh vertices distributions when faced with complex hand images.
\begin{figure}[htp]
\setlength{\abovecaptionskip}{2pt}
\setlength{\belowcaptionskip}{2pt}
  \centering
  \includegraphics[width=0.85\linewidth]{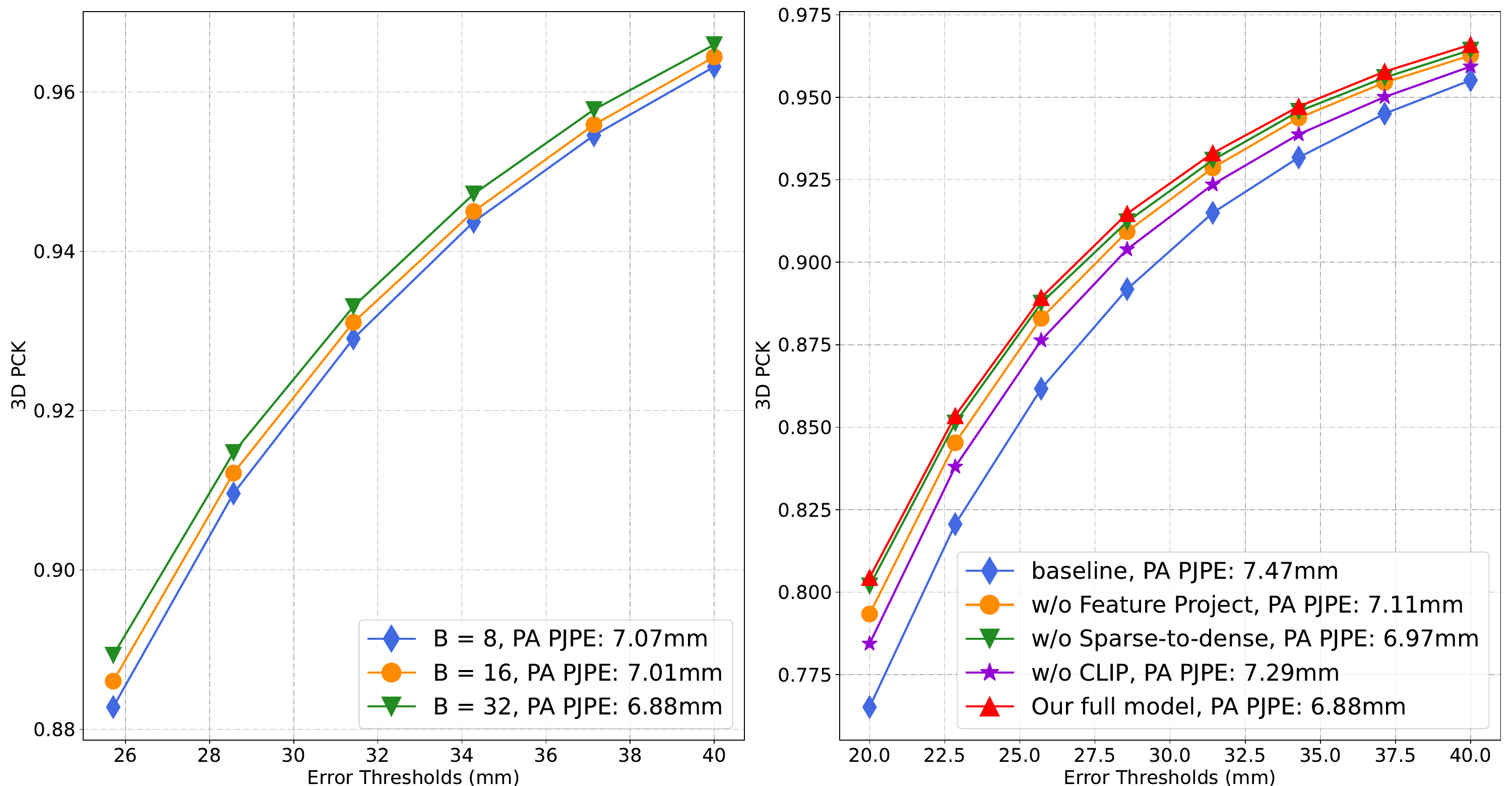}
  \caption{3D PCK curve comparison. 
  It measures the inference performance of network structures with different configurations on the FreiHAND validation set.}
  \label{ablation_auc}
\end{figure}
\begin{figure}[htp]
\setlength{\abovecaptionskip}{2pt}
\setlength{\belowcaptionskip}{2pt}
  \centering
  \includegraphics[width=0.85\linewidth]{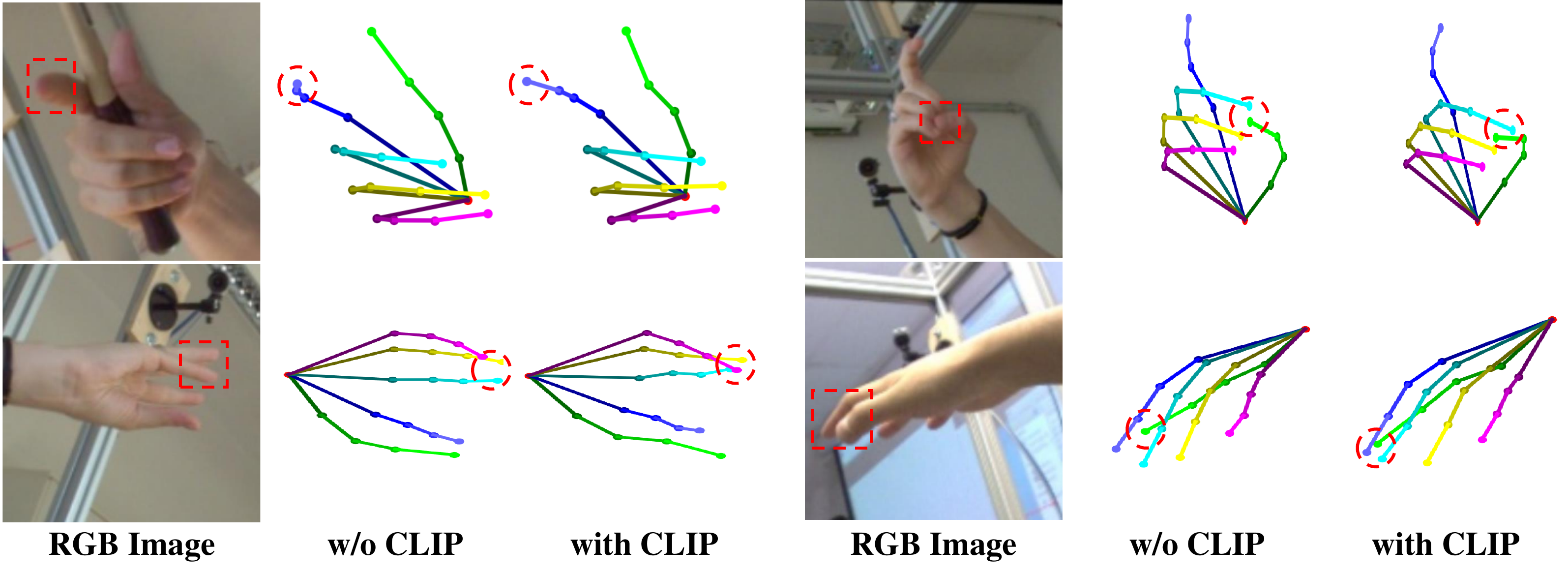}
  \caption{
  Ablation Study.
  Comparing the pose estimation results with or w/o CLIP module.
  }
  \label{ablationvis}
\end{figure}
Additionally, to further qualitatively demonstrate the role of CLIP, as shown in Fig.~\ref{batchvis}, we provide the connection matrices between text representations and visual features in a mini-batch containing 8 images. 
Whether in the "\textcolor[RGB]{65,105,225}{\textit{Left to Right}}" or "\textcolor[RGB]{255,127,36}{\textit{Top to Bottom}}" logit latent matrices, the regular distribution of highlighted values along the diagonal, which intuitively demonstrates that the proposed method can semantically connect text prompts and pose space distributions. 
Of course, besides the elements on the matrix diagonal, there are several "highlights" in the matrix, representing several text-image pairs in a mini-batch that conform to the general distribution in the respective direction.

\subsection{Ablation Study}
\textbf{Ablation Study of different structures:}
In Table.~\ref{ablation1}, we show the evaluation results of five methods with different model constraints on the FreiHAND validation set.
The most significant performance gap is between the method "w/o CLIP loss" and the "Full mode" (AUC: 0.763 vs. \textbf{0.776}, PA-PJPE: 7.29 vs. \textbf{6.88}); 
2) The method "w/o Feature Projection" is weaker than the "Full model" in all evaluation metrics (AUC: 0.769 vs. \textbf{0.776}, PA-PJPE: 7.11 vs. \textbf{6.88}); 
3) The performance of model "w/o Sparse-to-dense" structure (using a coarse-to-fine feature refinement strategy) is slightly behind the "Full model" (AUC: 0.773 vs. \textbf{0.776}, PA-PJPE: 6.97 vs. \textbf{6.88}); 
4) The "baseline" method without any sub-module or loss from 1), 2), and 3) has a significant gap compared to the "Full model" (AUC: 0.758 vs \textbf{0.776}, PA-PJPE: 7.47 vs \textbf{6.88}).
In summary, the supervision based on CLIP can enhance pose-aware representation by introducing text prompts, thereby improving model performance.
The proposed Feature Projection and designed Sparse-to-dense structure have a positive impact on model inference.

As shown in the right part of Fig.~\ref{ablation_auc}, we provide the 3D PCK curve within a given range and the corresponding PA-PJPE values to quantitatively demonstrate these results.
In Fig.~\ref{ablationvis}, we qualitatively compare the inference results of some images "w/o CLIP loss" and "with CLIP loss," using \textcolor{red}{red} dashed boxes to indicate image details and pose details separately. 
Taking the second image as an example, the estimation result supervised by CLIP loss can more accurately locate the position of the thumb fingertip on the left side of the middle fingertip.
\begin{table}[htp]
\setlength{\abovecaptionskip}{2pt}
\setlength{\belowcaptionskip}{2pt}
\caption{Ablation Study.
Comparison model performance on the FreiHAND validation set, under different batch size for CLIP-based learning.
}
\footnotesize
\setlength{\tabcolsep}{0.5mm}{
\begin{tabular}{@{}ccccc@{}}
\toprule
\diagbox{Batch Size}{Matrices}         & AUC(0-30mm)    & PJPE           & Median PJPE    & PA-PJPE       \\ \midrule
baseline & 0.763          & 14.86          & 12.27          & 7.29          \\
B = 8    & 0.770          & 14.44          & 11.96          & 7.07          \\
B = 16   & 0.772          & 14.24          & 11.87          & 7.01          \\
B = 32   & \textbf{0.776} & \textbf{13.95} & \textbf{11.58} & \textbf{6.88} \\
\bottomrule
\end{tabular}
}
\label{batchsize}
\end{table}

\textbf{Ablation Study of batch size in CLIP:}
To further clarify the impact of batch size on the proposed CLIP-based model, Table.~\ref{batchsize}, we demonstrate the influence of different batch sizes on inference performance for the FreiHAND validation set.
Compared to the "baseline" method without CLIP loss, inference accuracy positively correlates with the batch size \textbf{B}.
Specifically, when the "B = 16", the model performance has a slight difference compared to that of "B = 32" (AUC: 0.772 vs. \textbf{0.776}, PA-PJPE: 7.01 vs. \textbf{6.88}).
We analyze the reason that smaller batch sizes cannot effectively connect text prompts and pose-aware features due to it being unable to provide enough negative samples for the contrastive learning paradigm.
Regrettably, we cannot set a larger batch size to further verify the impact of this factor due to limited computing resources.
As depicted in the left part of Fig.~\ref{ablation_auc}, we provide the 3D PCK curve within a given range and the corresponding PA-PJPE values to quantitatively demonstrate the model performance under different batch size settings.
\begin{table}[htp]
\setlength{\abovecaptionskip}{2pt}
\setlength{\belowcaptionskip}{2pt}
\caption{Ablation Study.
Comparison model performance on the FreiHAND validation set under different number of select joints for text prompts generation.
}
\footnotesize
\setlength{\tabcolsep}{0.5mm}{
\begin{tabular}{@{}ccccc@{}}
\toprule
\diagbox{Configs}{Matrices}       & AUC(0-30mm) & PJPE  & Median PJPE & PA-PJPE \\ \midrule
N = 10 & 0.767       & 14.55 & 12.19       & 7.16    \\
N = 15 & 0.774       & 14.15 & 11.79       & 6.94    \\
\textbf{N = 21} & \textbf{0.776}       & \textbf{13.95} & \textbf{11.58}       & \textbf{6.88}    \\ \bottomrule
\end{tabular}
}
\label{selectjoints}
\end{table}

\textbf{Ablation Study of text prompts generation:}
In addition, we conducted ablation experiments to verify the effectiveness of different text prompt generation methods on the model.
In Table.~\ref{selectjoints}, we compared three methods' performance (10 joints, 15 joints, and 21 joints) of selecting hand joints and generating corresponding text prompts.
Compared to the method of selecting only 10 joints "N = 10" and converting them into text prompts, setting all 21 joints "N = 21" can encode a richer spatial distribution of hand joints, thereby guiding pose-aware features to more accurately locate hand joints detailed cues in 3D space.
\section{Conclusion}
This paper introduces CLIP-Hand3D, the first successful integration of text representations containing advanced human knowledge into 3D hand recovery. 
The proposed model achieves state-of-the-art performance on three public datasets, compared to methods employing similar-scale visual encoders, while significantly increasing the inference speed by a large margin ($\approx$28.3\%).
Specifically, The Text Feature Generation module converts the joint distribution order concealed within pose labels into text prompts and further matches pose-aware features with text representations through contrastive learning, improving the model performance by approximately 8.1\%. 
Additionally, the lightweight Mesh Regressor not only incorporates position encodings of varying scales but also queries joint-related semantic cues from the latent feature pyramid. 
We aim to delve deeper into the relationship between text and vision in the future for more flexible hand image understanding.



\bibliographystyle{ACM-Reference-Format}
\balance
\bibliography{sample-base}

\clearpage
\appendix
\section{APPENDIX}
\subsection{Inference Speed Analysis}
The inference process of the model is divided into two components: Pose Feature Generation and Mesh Regression. 
As depicted in Table~\ref{timecost}, we provide the inference speeds of these two sub-models on a single NVIDIA RTX 2080 Ti using the PyTorch framework. 
The Pose Feature Generation takes 6.59ms, while the Mesh Regressor takes 6.31ms, resulting in a total of 12.9ms, equivalent to approximately 77.5FPS.
Note that the Pose Feature Generation module encompasses several sub-modules: 
a Visual Encoder with a ResNet50 backbone, a Joint Regression based on the Lixel map module, and a Hand Joint Projection Layer. 
Besides, the Mesh Regressor module includes Heatmap Generation, a Multi-Transformer Layer, and several Multilayer Perceptron (MLP) layers.
\begin{table}[htb]
\caption{Inference speed of each module in our model.}
\footnotesize
\centering
\setlength{\tabcolsep}{0.45mm}{%
\begin{tabular}{@{}cccc@{}}
\toprule
{Module}               & {Pose Feature Generation} & {Mesh Regressor} & {Full Model} \\ \midrule
Inference Time(s)      & 0.00659                 & 0.00631        & 0.0129     \\ \midrule
Frames Per Second(FPS) & 151.7                   & 158.4          & 77.5       \\ \bottomrule
\end{tabular}%
}
\label{timecost}
\end{table}
\begin{figure}[ht]
  \centering
  \includegraphics[width=0.95\linewidth]{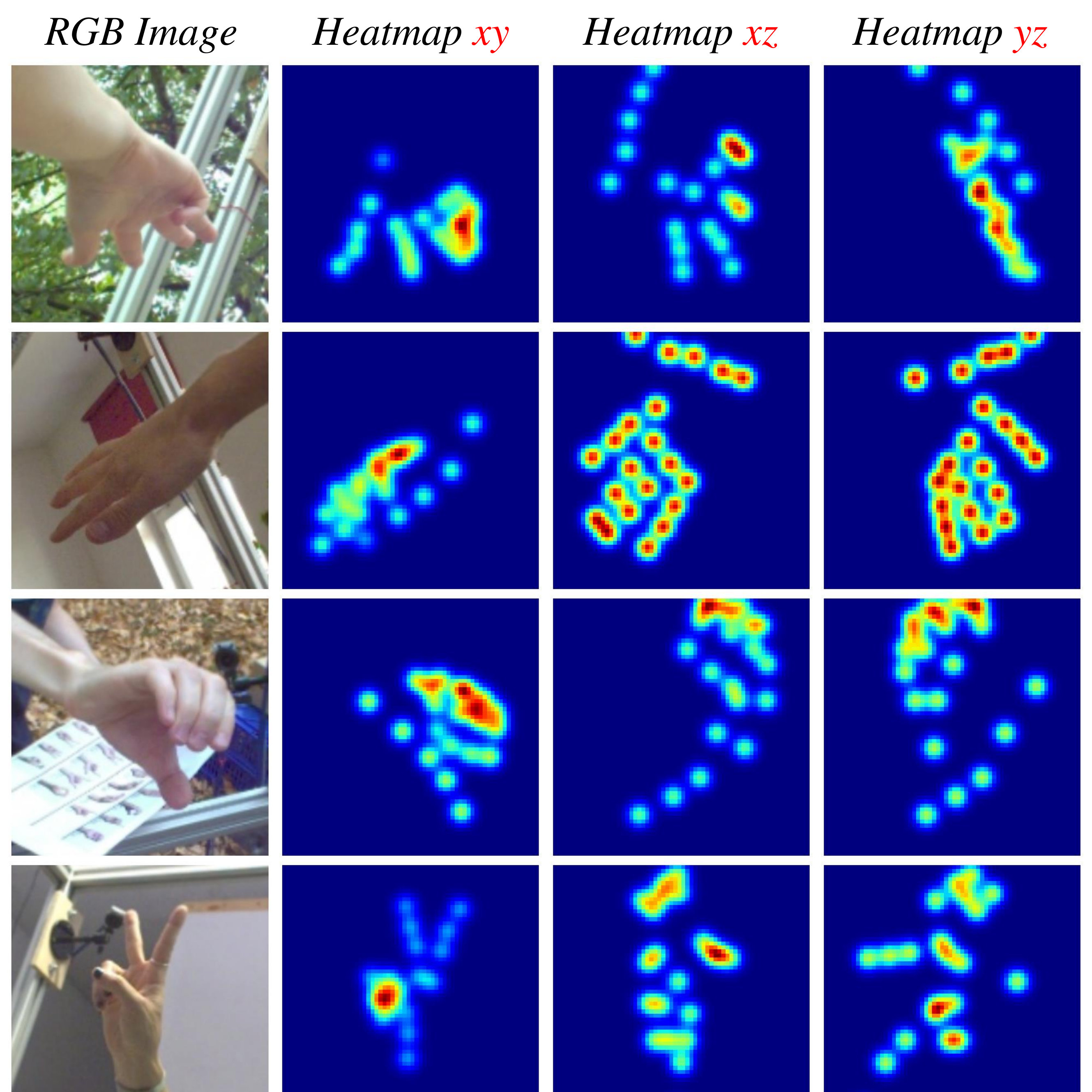}
  \caption{
  Qualitative evaluation. 
  The 2D heatmap distribution corresponding to the 21 hand joints in various dimensions is represented by heatmap xy $H_p^{xy}\in{R}^{(1, 56, 56)}$, heatmap xz $H_p^{xz}\in{R}^{(1, 56, 56)}$, and heatmap yz $H_p^{yz}\in{R}^{(1, 56, 56)}$.
  }
  \label{heatmap}
\end{figure}
\subsection{3D Joint Heatmap Analysis}
As we discussed in the main text, the objective of the Pose Feature Generation module is to accurately predict 3D joint positions within the \textit{uvd} space. 
However, sparse joint representation tends to dilute the rich semantic information inherent to the hand structure, given that discrete spatial points may not fully encapsulate the hand model. 
Therefore, mapping discrete joint positions to a 3D heatmap representation $H_p^{3D}\in{R}^{(21, 56, 56, 56)}$ is both a direct and effective approach.
To retain more rich semantics of the hand model, we choose to generate the corresponding 3D joint heatmap and pass it to the subsequent Mesh Regressor module.
Specifically, given the predicted joint positions in the \textit{uvd} space, the 2D heatmap in the xy, xz, and yz pixel planes encodes the position distribution of the hand model in the current orthogonal basis direction. 
As illustrated in Fig.~\ref{heatmap}, we provide the heatmap results for several images selected from the FreiHAND testset. 
These include the heatmap \textcolor{red}{xy} $H_p^{xy}\in{R}^{(21, 56, 56)}$, heatmap \textcolor{red}{xz} $H_p^{xz}\in{R}^{(21, 56, 56)}$, and heatmap \textcolor{red}{yz} $H_p^{yz}\in{R}^{(21, 56, 56)}$.
Formally, we have:
\begin{equation*}
    \mathcal{H}_i(u)=exp(-\frac{1}{2}||u-P_{i}^{3D}||^2/\sigma^2), u=[u_x, u_y, u_z],
\end{equation*}
where $u$ represents the joints index in the 3D \textit{uvd} space, $P_{i}^{3D}$ denotes the predicted 3D hand pose in the \textit{uvd} space, $\sigma$ = 2.0.
\subsection{Matching Matrix Analysis}
\begin{figure*}[]
  \includegraphics[width=0.75\textwidth]{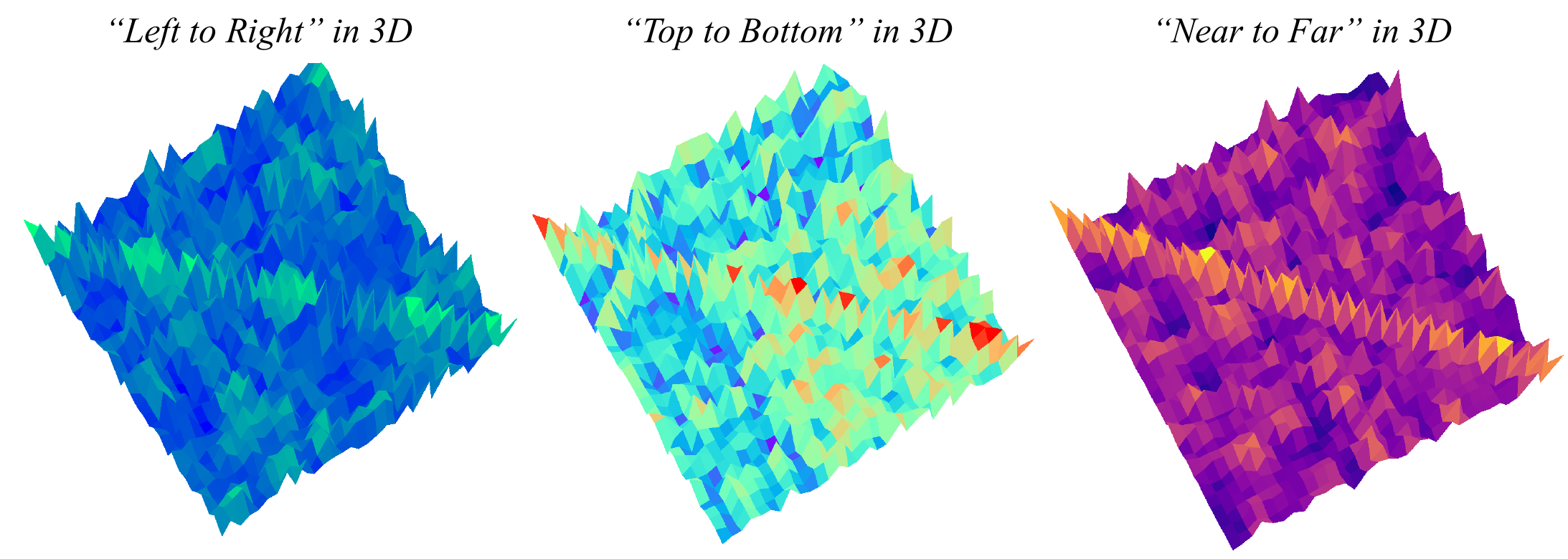}
  \caption{Qualitative Evaluation.
    We provide the 3D logits latent matrices corresponding to a batch of 32 samples, organized in the sequence of “Left to Right”, “Top to Bottom”, and “Near to Far”.}
  \label{3Dlogits_matrix}
\end{figure*}

As we discussed in the main text, the logits latent matrix serves as the bridge linking pose-aware features and text representation. 
To further illustrate this concept qualitatively, refer to Fig.~\ref{3Dlogits_matrix}, which displays the logits latent matrices (“Left to Right” $M_{lr}\in{R}^{(32, 32)}$, “Top to Bottom” $M_{tb}\in{R}^{(32, 32)}$, and “Near to Far” $M_{nf}\in{R}^{(32, 32)}$) that correspond to a batch of 32 samples during the model inference process. 
It should be noted that these logits latent matrices in 3D space provide a more intuitive visualization of the distribution of image-text match pairs, with the highest matching values predominantly distributed along the diagonal of the matrix.



\begin{figure*}
  \includegraphics[width=0.85\textwidth]{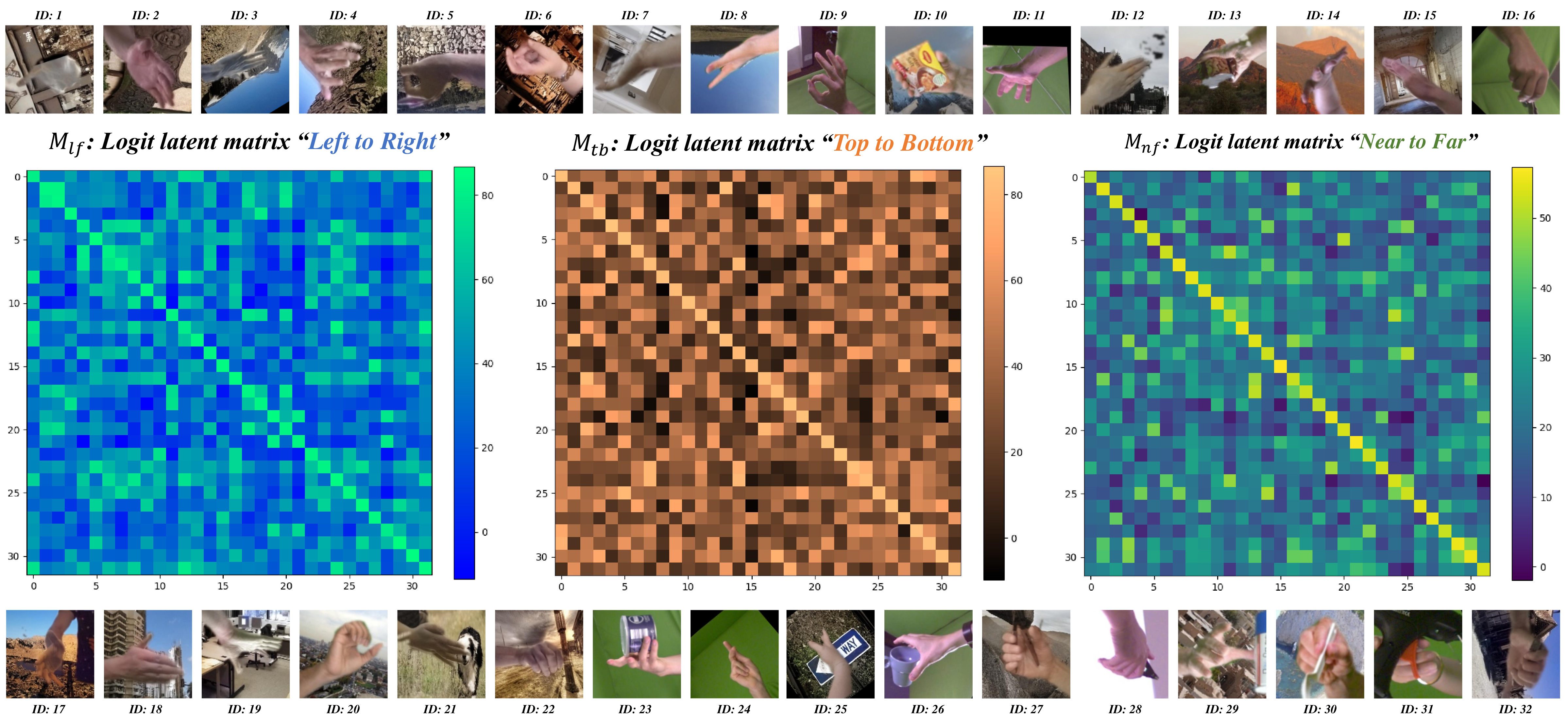}
  \caption{Qualitative Evaluation.
    We present the logits latent matrix corresponding to a batch of 32 samples, arranged in the sequence “Left to Right”, “Top to Bottom”, and “Near to Far”. 
    Besides, we provide a comprehensive visualization of all the images associated with this batch, each labeled with identification numbers ranging from 1 to 32.}
  \label{batch32}
\end{figure*}

To further increase the capacity of the logits latent matrix, as shown in Fig.~\ref{batch32}, the batch size has been further expanded to 32, and three matrices containing 32 image samples are displayed to qualitatively describe the image-text matching results.
From these illustrations, we can observe the following conclusions: 
1) the highest values are distributed on the diagonal to indicate that the pose-aware features and text representations have successfully paired; 
2) the matrix is symmetrically distributed along the diagonal, verifying the mutual matching between the text representations and pose-aware features; 
3) several localized highlights denote trends in similar pose distribution, further illuminating the effectiveness of our model.

\subsection{Limitations and Discussions}
\begin{figure}[htp]
  \centering
  \includegraphics[width=0.7\linewidth]{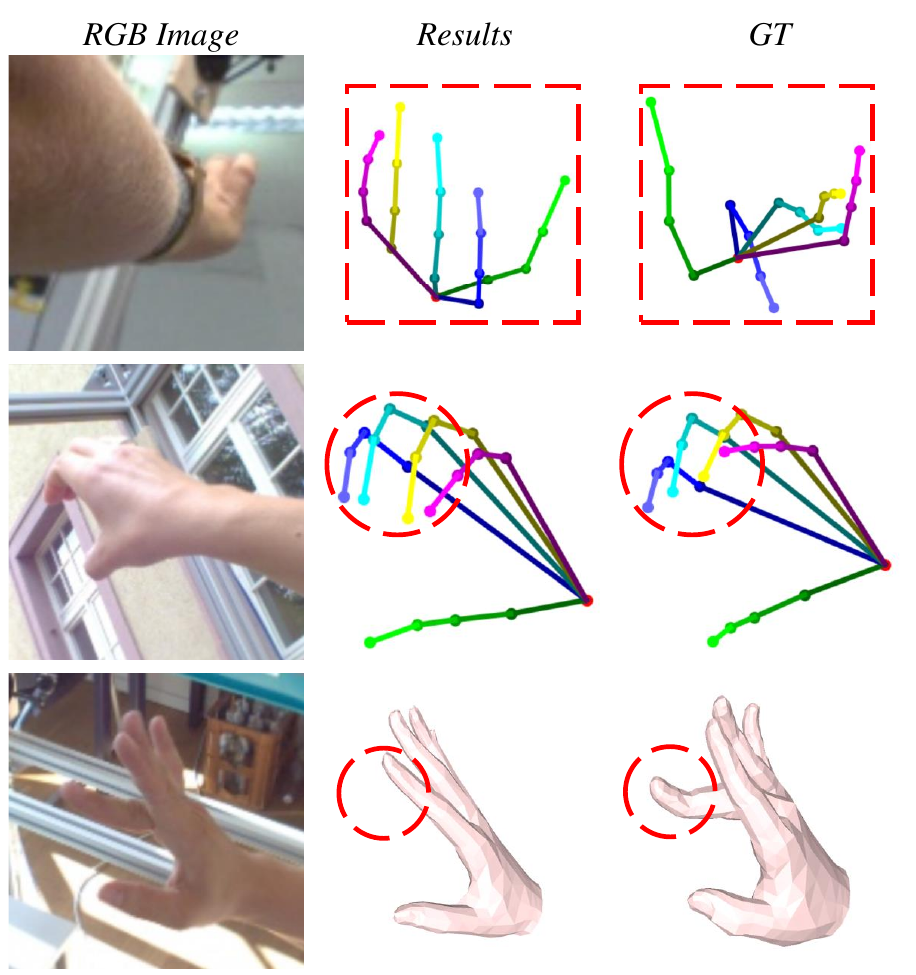}
  \caption{Failure inference cases.}
  \label{Failure cases}
\end{figure}

To further investigate the current limitations of the model, we have selected two representative failure inference cases for qualitative illustration. 
As depicted in the first row of Fig.~\ref{Failure cases}, the model deduces an unreliable 3D hand pose due to significant self-occlusion.
Our analysis suggests that, despite the integration of advanced human knowledge into the deep neural network via text prompts, the key to determining inference performance remains the pose-aware features.
Particularly in cases of implicit self-occlusion, the network is unable to provide accurate pose-aware encoding.

In addition, as shown in the second row of Fig.~\ref{Failure cases}, partially occluded hand semantics also cannot encode detailed joint distribution due to the lack of corresponding image texture, even with the assistance of text representation. 
As for the failure cases of 3D hand shape estimation, we show a representative case in the last row of Fig.~\ref{Failure cases}. 
In this case, due to the color texture of the middle finger being very similar to the background, the method mistakenly identifies it as the background, thereby providing incorrect cues to the subsequent pose-aware features, which indirectly affects the feature matching module.
In future work, we aim to incorporate text prompts more extensively to rectify the self-occlusion situation of hand images from the perspective of human high-level knowledge, thereby optimizing the joint distribution encoded in pose-aware features, that is, strengthening the role of text representation in the 3D hand recovery model.
\end{document}